\documentclass[12pt]{article}

\DeclareSymbolFont{UPM}{U}{eur}{m}{n}
\DeclareMathSymbol{\umu}{0}{UPM}{"16}
\let\oldumu=\umu
\renewcommand\umu{\ifmmode\oldumu\else$\oldumu$\fi}
\newcommand\micro{\umu}
\def\micron{\micro m}
\let\microns \micron
\let\oldmsp=\sp
\let\oldmsb=\sb
\def\sp#1{\ifmmode
	   \oldmsp{#1}%
	 \else\strut\raise.85ex\hbox{\scriptsize #1}\fi}
\def\sb#1{\ifmmode
	   \oldmsb{#1}%
	 \else\strut\raise-.54ex\hbox{\scriptsize #1}\fi}
\def\by{\ifmmode\times\else$\times$\fi}
\def\ttt#1{10\sp{#1}}
\def\tttt#1{\by\ttt{#1}}

\def\degree{\ifmmode\sp\circ\else$\sp{\circ}$\fi}
\let\degrees \degree

\usepackage{scicite}

\usepackage{times}

\usepackage{epsf}

\newenvironment{sciabstract}{%
\begin{quote} \bf}
{\end{quote}}

\newcounter{lastnote}
\newenvironment{scilastnote}{%
\setcounter{lastnote}{\value{enumiv}}%
\addtocounter{lastnote}{+1}%
\begin{list}%
{\arabic{lastnote}.}
{\setlength{\leftmargin}{.22in}}
{\setlength{\labelsep}{.5em}}}
{\end{list}}

\title{The Phase-Dependent Infra-Red \\ Brightness of the Extrasolar \\ Planet $\upsilon$~Andromeda~b}

\author{
Joseph Harrington,$^{1,2}$
Brad M. Hansen,$^{3\ast}$
Statia H.\ Luszcz,$^{2,4}$
Sara Seager,$^{5}$ \\
Drake Deming,$^{6}$
Kristen Menou,$^{7}$
James Cho,$^{8}$
L.\ Jeremy Richardson$^{9}$ \\
\\
\normalsize{$^{1}$Department of Physics, University of Central
Florida, Orlando, FL 32816, USA}\\
\normalsize{$^{2}$Center for Radiophysics and Space Research,} \\
\normalsize{Cornell University, Ithaca, NY 14853, USA}\\
\normalsize{$^{3}$Department of Physics and Astronomy
\& Institute of Geophysics and Planetary Physics}, \\
\normalsize{University of California at Los Angeles, Los Angeles, CA
90095, USA }\\
\normalsize{$^{4}$Department of Astronomy, University of California,
Berkeley, CA 94720, USA} \\
\normalsize{$^{5}$Department of Terrestrial Magnetism,} \\
\normalsize{Carnegie Institution of Washington, Washington DC 20015, USA} \\
\normalsize{$^{6}$Planetary Systems Laboratory and Goddard Center for
Astrobiology,} \\
\normalsize{ Goddard Space Flight Center, Greenbelt, MD 20771, USA}\\
\normalsize{$^{7}$Department of Astronomy, Columbia University, New
York, NY 10027, USA}\\
\normalsize{$^{8}$School of Mathematical Sciences,} \\
\normalsize{Queen Mary, University of London, London E1 4NS, UK}\\
\normalsize{$^{9}$Exoplanets and Stellar Astrophysics Laboratory,} \\
\normalsize{Goddard Space Flight Center, Greenbelt, MD 20771, USA} \\
\\
\normalsize{$^\ast$To whom correspondence should be addressed; E-mail:
hansen@astro.ucla.edu.}
}

\date{}

\begin{document} 




\maketitle


\begin{sciabstract}

The star $\upsilon$ Andromeda is orbited by three known planets\cite{But2},
the innermost of which has an orbital period of 4.617~days and a mass
 at least 0.69 that of Jupiter. This planet is close enough
to its host star that the radiation it absorbs 
 overwhelms its internal heat losses.
Here we present the 24-{\micron} light curve of this system, obtained
with the Spitzer Space Telescope.  It shows a clear variation in phase
with the orbital motion of the innermost planet.  This is the first
 demonstration that such planets possess
distinct hot substellar (day) and cold antistellar (night) faces.
\end{sciabstract}


Last year, two independent groups\cite{Dem1,Charb1} reported the
first measurements of the infrared light emitted by 
extrasolar planets orbiting close to their parent stars.  These `hot Jupiter'\cite{Vulcan} planets
have small enough orbits that the energy they absorb from their hosts
 dominates their own internal energy losses.
How they absorb and reradiate this energy is fundamental to 
understanding the behavior of their atmospheres.  One way to address
this question is to monitor the
emitted flux over the course of an orbit, to see whether the heat is
distributed asymmetrically 
about the surface of the planet. 

We have observed the $\upsilon$~Andromeda system with the 24-{\micron} channel of the Multiband Imaging Photometer for Spitzer (MIPS)\cite{Rieke}
aboard the Spitzer Space Telescope\cite{Werner}.  We took 168 three-second images 
 at each of five
epochs spread over 4.46 days (97\% of the 4.617-day orbital period of $\upsilon$~Andromeda~b)
beginning on 18~February 2006 at 12:52 UTC (Coordinated Universal Time).
After rejecting frames with bad pixels near the star and those with Spitzer's `first 
frame effect'\cite{Dem1} (2\% -- 8\% of the data, depending on epoch), we measure
the flux of the system and that of the surrounding sky using both sub-pixel, interpolated aperture photometry 
and optimal photometry\cite{Horne,Rich1} on each frame.

The detection of eclipses\cite{Sec} from the ``hot Jupiter'' planetary systems HD 209458b\cite{Dem1}, TrES-1\cite{Charb1} and
HD 189733b\cite{Dem2} demonstrate that a small fraction ($\sim 0.1\%$) of the total 
infrared light we observe from these systems is actually emitted from the planet rather than
the star. Thus, if we can measure the flux of a
system at a signal-to-noise ratio (S/N)
$>1000$, temperature differences between the day and night faces of the planet
will appear as an orbital modulation of the total system flux. With a star as bright as $\upsilon$~Andromeda,
our three-second exposures each have S/N $\sim 500$, so that our SNR
expectation is $\sim \sqrt{160} \times 500 \approx 6300$ at each epoch.

The MIPS instrument acquires data by placing the stellar image in a sequence of 
14 positions on the detector.  The detector's response varies
with position at the $\sim 1\%$ level.  This variation is stable and reproducible,
so we calculate correction factors as follows:  At each epoch we compute the
mean measured system flux at each position and take the ratio with the mean in
the first position.  We then average this ratio over all epochs for each position. This
results in corrections $< 2\%$ between positions, with uncertainties $\sim 6\times 10^{-4}$.
Bringing the photometry to a common normalization
allows us to average over all the frames in each epoch to
achieve S/N $\approx 4350$ at
each epoch.



As with most infrared instruments, MIPS's sensitivity varies in time.  
We correct for such drifts by 
dividing the system flux value by the measured background in each
frame.  The background at 24 {\microns} is thermal emission from
the zodiacal dust.
This dust pervades the inner solar system, absorbing light
from the sun and reradiating it at infrared wavelengths.  At 24 {\microns}, its emission is strong
enough for use as a flux standard, a technique used successfully
in measuring the eclipse of HD 209458b\cite{Dem1}.
However, the present work requires one additional correction.
The zodiacal background is the integrated emission by dust along the line of sight between the telescope and
the object.  The observed value thus
undergoes an annual modulation as that line of sight varies with the
telescope's orbit about the sun.  
The best available model\cite{DIRBE} predicts a linear drift over the brief
interval of our observations.  However, we cannot use the Spitzer model directly, since it is calculated for a line of
sight from the Earth to the object in question.  The difference in position between the Earth-trailing
telescope and the Earth itself is large enough that the slope of the
variation may be slightly different. Thus, we fit for the linear drift directly, simultaneously
with any model lightcurve fits.

\begin{figure}
\epsfxsize=6truein
\epsfbox{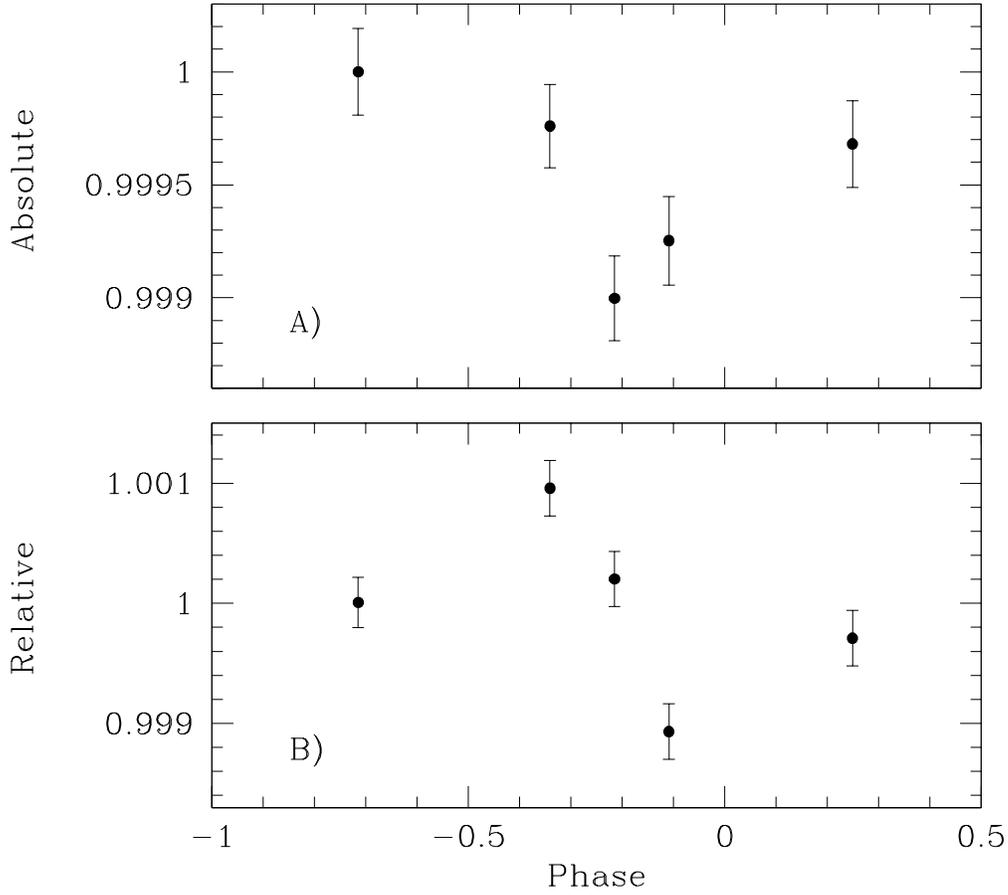}
\caption{The light curve of the $\upsilon$~Andromeda system.
A) The phase variation in the $\upsilon$~Andromeda system flux
before any corrections are applied for instrument or zodiacal drifts. Variations in the system flux are significant
even at this point.
B) By comparing to the zodiacal background, and
fitting for the linear drift in the background due
to the telescope's motion, we obtain the phase curve shown.
In each case, phase is shown modulo unity, with zero phase occurring when
the planet is closest to earth. The amplitude units are expressed in terms of the system flux at the first
epoch.
 \label{Panel}}
\end{figure}

The phase curve for the
$\upsilon$~Andromeda system shows a variation (Fig~\ref{Panel}) in absolute photometry, even before any
corrections for instrumental or zodiacal drifts are made.
After the calibration with respect to the zodiacal background is applied, this variation is revealed to be
 in phase with the known orbit of the innermost
planet of the system, our principal result. 

\begin{figure}
\epsfxsize=6truein
\epsfbox{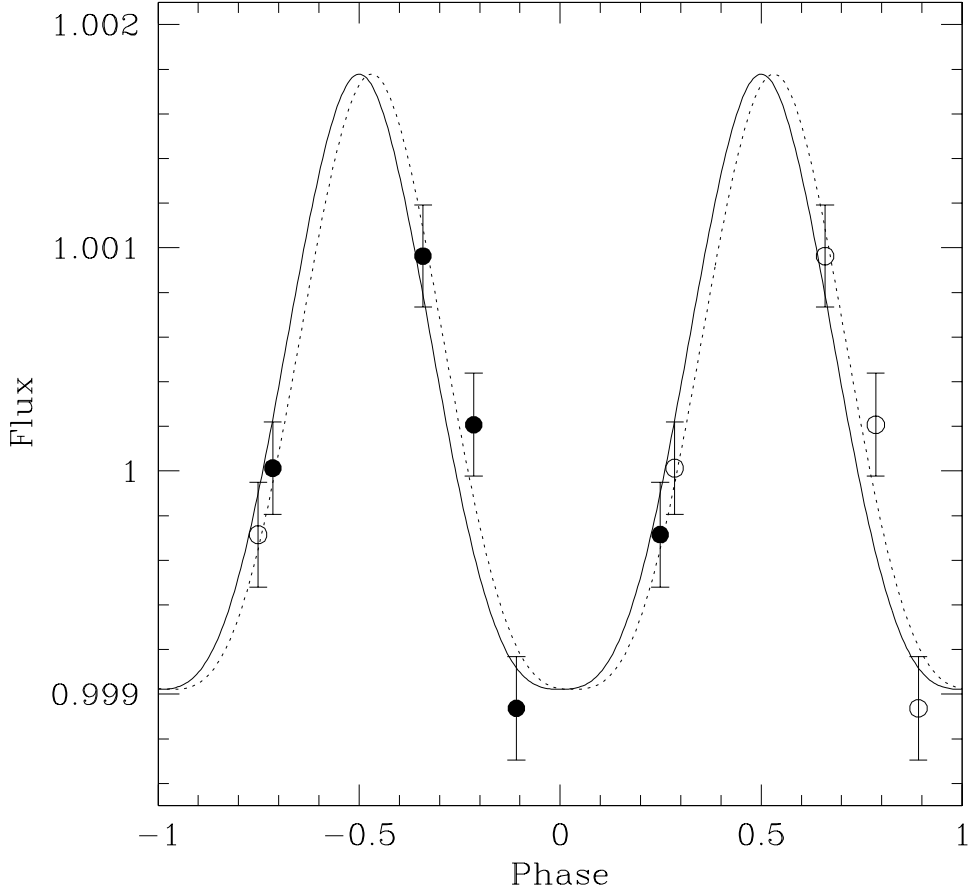}
\caption{Comparison of the phase curve and the No-Redistribution Model.
The solid points show our final phase curve, after applying calibrations, in time
order from left to right. The open points are repetitions of these, displaced horizontally by one orbit,
to better illustrate the phase coverage over two cycles.
The solid line is an analytic model for the planetary
emission in which energy absorbed from the star is reradiated locally on the day side with no heat
transfer across the surface of the planet, the
so-called No-Redistribution model (and in excellent agreement with the more detailed version in
 \cite{Bar}). The assumed inclination in this case is 80{\degrees} from pole-on, and the
relative planet/star amplitude is $2.9\times 10^{-3}$.
If we allow for a phase shift relative to the radial velocity curve, we  obtain a slightly better
fit, as shown by the dotted curve. The best fit is obtained with a phase lag of $11^{\circ}$, but
zero lag is excluded only at the 2.5$\sigma$ level. \label{Fit}}
\end{figure}

A simple model can be fit to the phase curve (Fig~\ref{Fit}), assuming
local, instantaneous thermal reradiation of the absorbed stellar flux.
In the simplest model, the phase of the variation is not a free parameter, but 
 is rather set by the measured radial velocity curve\cite{But1}.
Phase offsets are possible for
 models in which the energy is absorbed deep within the atmosphere 
and redistributed about the surface\cite{Cho,CS}. There is weak (2.5$\sigma$) evidence
for a small phase offset in this data (Fig~\ref{Fit}) but the large offsets
predicted from some models are excluded at high significance.
Fitting the
peak-to-trough amplitude to the observations yields a best-fit value for the planet-star flux ratio 
$2.9 \pm 0.7\times 10^{-3}$ times the star's brightness.  This is very similar to
the result at this wavelength for 
HD~209458b\cite{Dem1}.  However, the latter is a measure of the 
absolute flux from the planet divided by that from the host star, while
the present result 
is a measure of the flux difference between the projected day and night sides, 
divided by the flux of the (different) host star.

\begin{figure}
\epsfxsize=6truein
\epsfbox{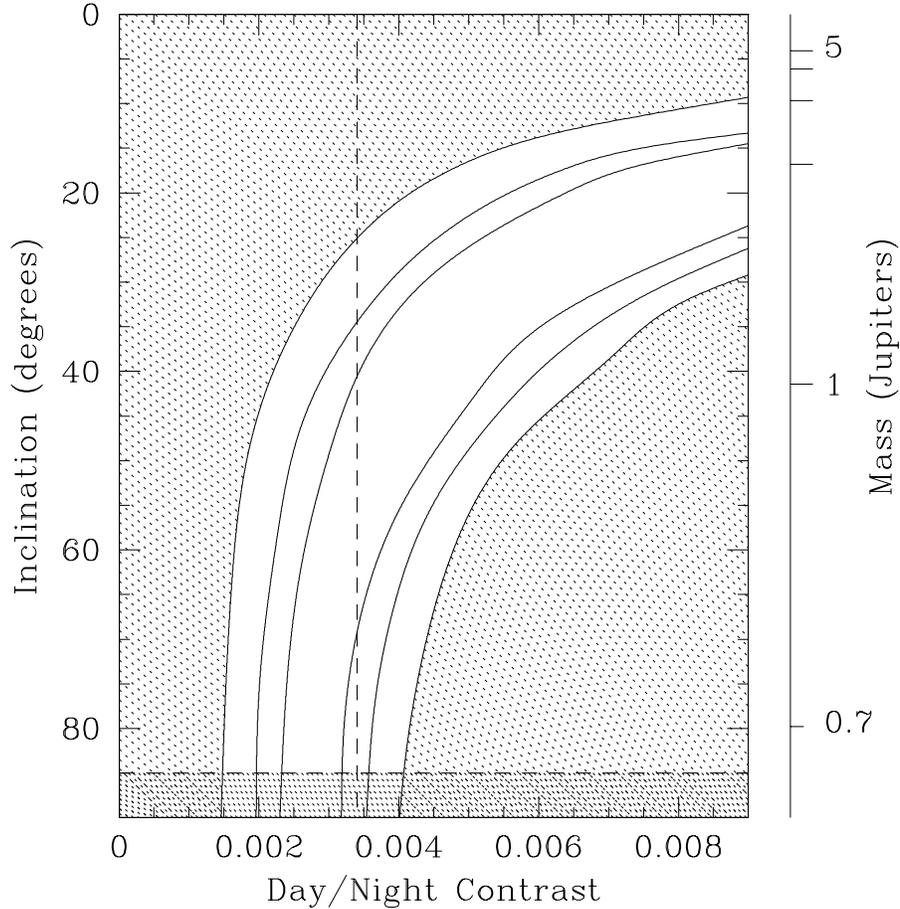}
\caption{The influence of inclination on the inferred Day-Night contrast.
The solid contours bound the $1, 2$ and $3\sigma$ confidence regions for the day-night flux difference (in
units of the stellar flux), determined as a function of assumed orbital inclination (measured
relative to a face-on orbit). The large shaded regions indicate those values excluded at 3 $\sigma$.
The lower shaded region is excluded because the planet does not transit in front of
the star. The vertical dashed line indicates the expected upper limit to the contrast, obtained when the 
night side is completely dark and all of the stellar flux is reradiated from the day side,
in accordance with the no-redistribution model, and assuming zero albedo. 
At the right we show the true mass of the planet given the assumed
inclination (based on the minimum mass derived from the radial velocity curve), in units of Jupiter masses.
 \label{IC}}
\end{figure}

Another difference between the cases of $\upsilon$~Andromeda~b and HD 209458b is that we do not have a strong
constraint on the orbital inclination in this system, so we must include the unknown inclination in the
 model fit (Fig~\ref{IC}).
At higher inclinations, parts of both the night side and the day side
are always visible,
so the true contrast between the day and night sides must be larger
than the amplitude of the observed variation.
This contrast is ultimately driven by the light absorbed from the
star, which therefore provides
an upper limit.
We know the distance of the planet from the star and the stellar properties, so we can estimate the
contrast that would result if all of the observed flux were reradiated from the day side and nothing from the night side.
If we assume the planet's radius is $<$1.4 Jupiter radii (as observed for other planets of this class), then we can
constrain the expected amplitude to be $<3.4\times 10^{-3}$ (2$\sigma$) for a simple black-body, no-redistribution model
with zero albedo. Thus, a consistent picture of the atmospheric energetics emerges as long as the orbital inclination
is $>$30{\degrees}.


A natural question to ask is whether there are any plausible alternative models for the observed variation. 
 The estimated rotation period of
the star is too long to explain our phase curve as the result of a normal starspot (which is darker than
other parts of the stellar surface).
One could posit a feature on the stellar surface similar to a starspot but 
induced by a magnetic interaction between the star and the planet, and therefore moving synchronously with the planet.
However, \cite{Bali} place an upper limit of $1.6 \times 10^{-4}$ on the amplitude of optical variation with the
planetary orbital period, so infrared variability from the star should be even weaker than this. Some evidence
for such magnetospheric interactions is found in observations
of chromospheric calcium H and K lines\cite{Shkol} and has even been seen in the $\upsilon$~Andromeda system. 
However, the energy input needed to explain the Ca lines is
$\sim10^{27}\rm  ergs \, s^{-1}$, much less than the minimum planetary luminosity we
infer here
 ($\sim 4\times 10^{29} \rm ergs\, s^{-1}$).
Indeed, one can make a quite general argument that our observations
cannot be powered by the same mechanism, since any
heating of the star due to magnetic interaction with the planet
ultimately extracts energy from the planetary orbit.  Thus, one
may calculate an orbital decay time 
$$
\tau = \frac{G M_* M_p}{2 a \dot{E}} = 5\tttt{6} {\rm yr} \left( \frac{M_p}{M_J} \right)
\left( \frac{a}{12 R_{\odot}} \right)^{-1} \left( \frac{\dot{E}}{\rm 10^{30} ergs\; s^{-1}} \right)^{-1},
$$
where $M_*$ and $M_p$ are the stellar and planetary masses, $a$ is the semi-major axis and $\dot{E}$ is
the observed heating rate.
Heating at the level necessary to explain our observations would result in the decay of the 
planetary orbit on timescales $<$\ttt{7}~years, while the estimated age of the system is 3~Gyr.
As such, the chromospheric heating of the star is unlikely
to be related to the effect seen at 24 {\microns}.

\begin{figure}
\epsfxsize=6truein
\epsfbox{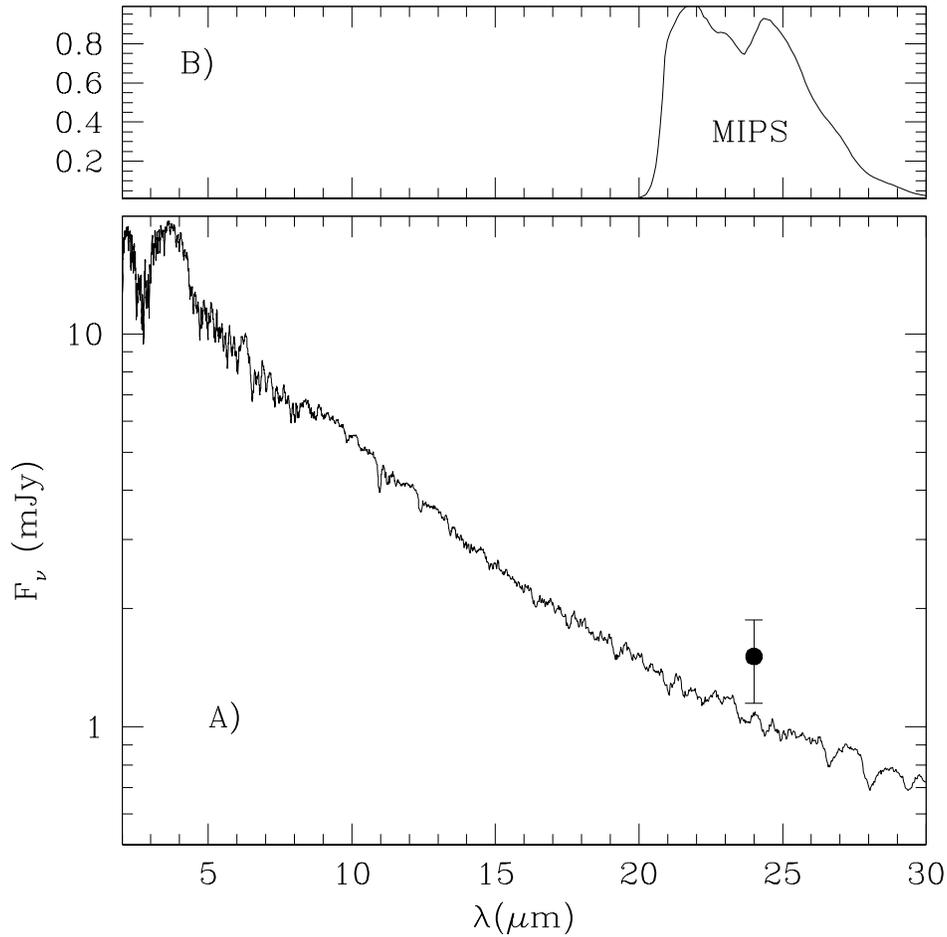}
\caption{Comparison of the measured amplitude and a planetary spectral model.
A) The solid curve shown is a model\cite{Seag} for a planet of radius 1.4~$R_{\rm J}$, irradiated with parameters appropriate
to the $\upsilon$~Andromeda system observed at full phase. This results in a temperature
$\sim 1977$~K\cite{Teq}. 
The model is in agreement with the observations (filled circle)
at the $2\sigma$ level. B) The normalised spectral response curve of the MIPS 24 $\mu$m instrument extends from
20$\mu$m to 30$\mu$m. 
\label{Spec2}}
\end{figure}

This observation  reveals the presence of a temperature asymmetry on the surface
of an extrasolar planet.  The first measurements of eclipses\cite{Dem1,Charb1}
yielded measurements of the absolute flux levels emerging from the day sides of two extrasolar
planets.  When compared with models of radiative transfer in such atmospheres\cite{Seag,Bar,Fort,Burr},
those observations are consistent with a situation intermediate between no redistribution and full redistribution.  
A similar comparison is possible in this case (Fig~\ref{Spec2}). Our observed day-night flux difference is comparable
to the flux emerging at full phase in the models of \cite{Seag}, which indicates that there is little evidence for
redistribution of energy to the night side.

In conclusion, the observation of the phase curve of $\upsilon$~Andromeda~b indicates that significant temperature
differences exist between the day and night faces of the planet, consistent with a model in which very little horizontal 
energy transport occurs in the planetary atmosphere.  Furthermore, it indicates that the opportunities for direct extrasolar planetary
observations are better than previously thought, since useful data can
be obtained even in cases where the planetary orbit is not so
fortuitously aligned that the system exhibits transits or eclipses.



\begin{scilastnote}
\item
This work is based on observations made with the Spitzer Space Telescope, which is operated by the
Jet Propulsion Laboratory, California Institute of Technology, under contract to NASA. Support for
this work was provided directly by NASA, its Origins of Solar Systems and Astrophysical Theory Programs as well as the 
Astrobiology Institute and Spitzer.  We thank the personnel of the Spitzer Science Center 
and its MIPS
instrument, who ultimately made these measurements possible.
\end{scilastnote}

\begin{thebibliography}{99}
\bibitem{Vulcan} There is no official terminology for planets that orbit close to their parent stars. The term ``Hot Jupiter''
is the most common, although some authors have adopted other terms, such as ``Pegasids'',``Pegasi planets'', or ``Roasters''.
\bibitem{Sec} The passage of an extrasolar planet behind its star ---
an eclipse --- results in a
drop in the total system flux.  This allows a measurement of the absolute flux emerging from 
the part of the planet that faces the star (the day side).  This is possible when the  orbit is aligned
almost edge-on to the line of sight.
\end{thebibliography}
\end{document}